\documentclass[usegraphicx,usenatbib]{mn2e}
\usepackage{amsmath}
\usepackage{amssymb}
\usepackage{color}
\usepackage{url}
\usepackage{verbatim}
\usepackage{threeparttable}

\newcommand\lsim{\mathrel{\rlap{\lower4pt\hbox{\hskip1pt$\sim$}}
        \raise1pt\hbox{$<$}}}
\newcommand\gsim{\mathrel{\rlap{\lower4pt\hbox{\hskip1pt$\sim$}}
        \raise1pt\hbox{$>$}}}
\newcommand{\lya}{\ifmmode\mathrm{Ly}\alpha\else{}Ly$\alpha$\fi}
\newcommand{\lyb}{\ifmmode\mathrm{Ly}\beta\else{}Ly$\beta$\fi}
\newcommand{\igm}{\ifmmode\mathrm{IGM}\else{}IGM\fi}
\newcommand{\lae}{\ifmmode\mathrm{LAE}\else{}LAE\fi}
\newcommand{\h}{\ifmmode\mathrm{H}\else{}H\fi}
\newcommand{\hi}{\ifmmode\mathrm{H\,{\scriptscriptstyle I}}\else{}H\,{\scriptsize I}\fi}
\newcommand{\hii}{\ifmmode\mathrm{H\,{\scriptscriptstyle II}}\else{}H\,{\scriptsize II}\fi}
\newcommand{\heii}{\ifmmode\mathrm{He\,{\scriptscriptstyle II}}\else{}He\,{\scriptsize II}\fi}
\newcommand{\cmb}{\ifmmode\mathrm{CMB}\else{}CMB\fi}
\newcommand{\qso}{\ifmmode\mathrm{QSO}\else{}QSO\fi}
\newcommand{\eor}{\ifmmode\mathrm{EoR}\else{}EoR\fi}
\newcommand{\cmmc}{\textsc{\small 21CMMC}}
\newcommand{\cmfst}{\textsc{\small 21CMFAST}}
\newcommand{\sense}{{\small 21}\textsc{cmsense}}
\newcommand{\CH}{\textsc{cosmohammer}}
\newcommand{\emcee}{\textsc{emcee}}
\newcommand{\fg}{\textsc{faint galaxies}}
\newcommand{\bg}{\textsc{bright galaxies}}

\title[21CMMC with a 3D light-cone]{21CMMC with a 3D light-cone: the impact of the co-evolution approximation on the astrophysics of reionisation and cosmic dawn.}

\author[B. Greig \& A. Mesinger] {Bradley~Greig$^{1,2,3}$\thanks{E-mail:~greigb@unimelb.edu.au} \& Andrei~Mesinger$^{3}$ \\
$^1$ARC Centre of Excellence for All-Sky Astrophysics in 3 Dimensions (ASTRO 3D), University of Melbourne, VIC 3010, Australia \\
$^2$School of Physics, University of Melbourne, Parkville, VIC 3010, Australia \\
$^3$Scuola Normale Superiore, Piazza dei Cavalieri 7, I-56126 Pisa, Italy
}

\begin{document}
\maketitle \begin{abstract}
\noindent
We extend \cmmc{}, a Monte Carlo Markov Chain sampler of 3D reionisation simulations, 
to perform parameter estimation directly on 3D light-cones of the cosmic 21cm signal.
This brings theoretical analysis closer to the tomographic 21-cm observations achievable with next generation interferometers like HERA and the SKA.
Parameter recovery can therefore account for modes which evolve with redshift/frequency.  Additionally, simulated data can be more easily corrupted to resemble real data.
Using the light-cone version of \cmmc{}, we quantify the biases in the recovered astrophysical parameters if we use the 21cm power spectrum from the co-evolution
approximation to fit a 3D light-cone mock observation.
While ignoring the light-cone effect under most assumptions will not significantly bias the recovered astrophysical parameters, it can lead to an underestimation of the associated uncertainty.
However significant biases  ($\sim$few -- 10 $\sigma$) can occur if the 21cm signal evolves rapidly (i.e. the epochs of reionisation and heating overlap significantly)
and:
(i) foreground removal is very efficient, allowing large physical scales ($k\lsim0.1$~Mpc$^{-1}$) to be used in the analysis or
(ii) theoretical modelling is accurate to within $\sim10$ per cent in the power spectrum amplitude.
\end{abstract} 
\begin{keywords}
cosmology: theory -- dark ages, reionisation, first stars -- diffuse radiation -- early Universe -- galaxies: high-redshift -- intergalactic medium
\end{keywords}

\section{Introduction}

The ubiquity of neutral hydrogen in the early Universe renders the intergalactic medium (IGM) completely opaque to
visible and ionising radiation. The formation of the first stars and galaxies, referred to as the Cosmic Dawn (CD), is thus 
enshrouded from even our most powerful optical and IR space-based telescopes, the {\it Hubble Space Telescope (HST)} and the 
forthcoming {\it James Webb Space Telescope (JWST)}. Over time, stars and galaxies begin to 
form more readily, increasing the cumulative amount of ionising radiation to the point that they are 
capable of producing local ionised (\hii{}) regions within the IGM. Eventually, these \hii{} regions percolate, resulting in
an ionised IGM. This corresponds to the final major baryonic phase change in the Universe, and is referred to as
the epoch of reionisation (EoR).

The most promising observable for investigating the EoR and CD is the 21 cm spin-flip 
transition of neutral hydrogen \citep[see e.g.][]{Gnedin:1997p4494,Madau:1997p4479,Shaver:1999p4549,Tozzi:2000p4510,
Gnedin:2004p4481,Furlanetto:2006p209,Morales:2010p1274,Pritchard:2012p2958}. This signal, in the commonly
assumed optically thin limit, is typically expressed 
as the offset of the 21cm brightness temperature, $\delta T_{\rm b}(\nu)$, relative to the temperature of the 
cosmic microwave background (CMB), $T_{\rm CMB}$ \citep[e.g.][]{Furlanetto:2006p209}:
\begin{eqnarray} \label{eq:21cmTb}
\delta T_{\rm b}(\nu) &\approx& 27x_{\hi{}}(1+\delta_{\rm nl})\left(\frac{H}{{\rm d}v_{\rm r}/{\rm d}r+H}\right)
\left(1 - \frac{T_{\rm CMB}}{T_{\rm S}}\right) \nonumber \\
& & \times \left(\frac{1+z}{10}\frac{0.15}{\Omega_{\rm m}h^{2}}\right)^{1/2}
\left(\frac{\Omega_{\rm b}h^{2}}{0.023}\right)~{\rm mK},
\end{eqnarray}
where $x_{\hi{}}$ is the neutral fraction, $T_{\rm S}$ is the gas spin temperature, 
$\delta_{\rm nl} \equiv \rho/\bar{\rho} - 1$ is the gas overdensity, $H(z)$ is the Hubble parameter, 
${\rm d}v_{\rm r}/{\rm d}r$ is the gradient of the line-of-sight component of the velocity and all quantities 
are evaluated at redshift $z = \nu_{0}/\nu - 1$, where $\nu_{0}$ is the 21 cm frequency. For brevity, we
drop the spatial dependence of the signal.

The spatial and frequency (redshift) dependence of the 21cm signal will
reveal a full three dimensional image of the IGM. It is sensitive to the ionisation
state (through $x_{\hi{}}$ in Equation~\ref{eq:21cmTb}), enabling an exploration of the EoR morphology and relevant sources.
It is also sensitive to the thermal state (e.g.\,$T_{\rm S}$), which allows us to study the high-energy processes in the first galaxies, thought to dominate the early heating of the IGM.  Empirical scaling relations based on the X-ray to star formation rates of nearby star-forming galaxies suggest that this epoch of heating (EoH) preceded the EoR (e.g. \citealt{McQuinn:2012p3773} and references therein).

This utility of the 21cm signal at exploring the early Universe has led to the 
construction and development of several experiments aiming to provide the first detection.
Broadly, these fall into two categories: (i) large-scale interferometric experiments measuring the
spatial fluctuations in the 21cm signal and (ii) global all-sky averaged experiments. In this work, we
shall focus on the former, which include: the Low Frequency Array 
(LOFAR; \citealt{vanHaarlem:2013p200,Yatawatta:2013p2980}), the Murchison Wide Field Array 
(MWA; \citealt{Tingay:2013p2997}), the Precision Array for Probing the Epoch of Reionisation 
(PAPER; \citealt{Parsons:2010p3000}), the Square Kilometre Array (SKA; \citealt{Mellema:2013p2975}) 
and the Hydrogen Epoch of Reionisation Array (HERA; \citealt{DeBoer:2017p6740}).

These experiments generate an enormous volume of data, sampling the full EoR/CD light-cone, with fields of view of thousands of square degrees.  In response to this upcoming Big-Data revolution, an enormous amount of effort has gone into theoretical interpretation.  The field of astrophysical parameter recovery has emerged, in analogy to the established CMB framework for physical cosmology.
Parameter recovery techniques include Monte Carlo Markov Chain (MCMC) on-the-fly sampling of 3D EoR/CD simulations \citep{Greig:2015p3675,Greig:2017p8496}, MCMC sampling emulators trained to reproduce summary statistics of 3D simulations \citep[e.g.][]{Kern:2017p8205,Schmit:2017}, machine learning on summary statistics \citep{Shimabukuro:2017p7140} and interpolating simulations over a parameter grid \citep{Mesinger:2014p244,Pober:2014p35,Hassan:2017}.

All of these analyses adopt the common co-evolution (`co-eval') approximation: each simulation box is assumed to have the same redshift along the line-of-sight. 
This is despite the fact that the 21cm signal varies along the line-of-sight owing to its frequency dependence, in effect producing an observational light-cone. 

The impact of the light-cone effect has been explored by several authors on either the correlation function \citep{Barkana:2006p8432,Zawada:2014p8433} 
or the 21cm power spectrum \citep[PS;][]{Datta:2012p7679,Datta:2014p4990,LaPlante:2014p7651,Ghara:2015p7650,Mondal:2017p8569}. 
Evolution of the cosmic signal along the light-cone is understandably most noticeable on large physical scales.
For example, the amplitude of the PS on large scales can be higher/lower in the initial/late stages of the EoR, compared to co-eval power spectra computed at the mean redshift over the band.
The overall strength of this behaviour can be mitigated by limiting the observing bandwidth of the experiment,
minimising the line-of-sight evolution by breaking the 21cm signal up into an increasing number of observations with smaller bandwidth. 
As a consequence though, the total number of large scale modes, which are the most sensitive to the astrophysics, are decreased. 
Though it is clear that the light-cone effect has an impact on the 21cm signal, it remains to be seen how significant a bias neglecting
it can have on the recovered astrophysical parameters.

In this work we extend \cmmc{}\footnote{https://github.com/BradGreig/21CMMC} \citep{Greig:2015p3675,Greig:2017p8496}, a Monte Carlo Markov Chain sampler of 3D EoR simulations,  to perform parameter estimation directly on 3D light-cones of the cosmic 21cm signal\footnote{To be precise, our simulations produce light {\it cuboids}, as the transverse edges are of constant co-moving scale, and not constant angular scale.  However, we follow the trend in the literature, and refer to our outputs as light {\it cones}.}.
To quantify  the impact of the light-cone effect, we consider two different reionisation histories, the \fg{} and \bg{}
models of \citet{Greig:2017p8496} and compare the recovery of the input astrophysical parameters with and without the light-cone effect. 
These two models constitute fast and slow reionisation histories, consistent with existing observational constraints.
We expect these two extremes to bracket the bias when performing parameter recovery using co-eval cubes.

Further, as we shift to mimic the observational signal as closely as possible, we 
also include a treatment of redshift space distortions (RSDs) along the line-of-sight. Sourced by the peculiar motions of the gas, the 
observed 21cm signal can be enhanced (suppressed) in the vicinity of large (small) density structures \citep[e.g.][]{Mao:2012p7838}.

The remainder of this paper is organised as follows.
In Section~\ref{sec:theory} we summarise the theoretical framework used to compute the cosmic 21-cm light-cones.
In Section \ref{sec:obs}, we present our mock observations, including the mock signal and telescope noise.
In Section \ref{sec:21cmmc} we summarise \cmmc{} before showing the parameter constraints obtained using the light-cones in Section~\ref{sec:Forecasts}.
We then quantify the corresponding bias in parameter recovery using co-eval simulations instead of the light-cone in Section~\ref{sec:bias} before finishing with our closing remarks in Section~\ref{sec:summary}.
Unless stated otherwise, we quote all quantities in co-moving units and adopt the cosmological parameters: 
($\Omega_\Lambda$, $\Omega_{\rm M}$, $\Omega_b$, $n$, $\sigma_8$, $H_0$) = (0.69, 0.31, 0.048, 0.97, 0.81, 68 km s$^{-1}$ Mpc$^{-1}$), 
consistent with recent results from the Planck mission \citep{PlanckCollaboration:2016p7780}.

%%%%%%%%%%%%%%%%%%%%%%%%%%%%%%%%%%%%%%%%%%%%%%%%%%%%%%%%%%%%%%%%%%%%%%%%%%%%%%%%%%%%%%%%%%%%%%%%%%%%%%%%%%%%%%%%%%%%%%%%%%%%%%%%
%%%%%%%%%%%%%%%%%%%%%%%%%%%%%%%%%%%  SIMULATING A LIGHTCONE  %%%%%%%%%%%%%%%%%%%%%%%%%%%%%%%%%%%%%%%%%%%%%%%%%%%%%%%%%%%%%%%%%%%
%%%%%%%%%%%%%%%%%%%%%%%%%%%%%%%%%%%%%%%%%%%%%%%%%%%%%%%%%%%%%%%%%%%%%%%%%%%%%%%%%%%%%%%%%%%%%%%%%%%%%%%%%%%%%%%%%%%%%%%%%%%%%%%%
\section{Calculating the 21cm light-cone}
\label{sec:theory}

\subsection{The ionisation and thermal state of the IGM} \label{sec:modelling}

We compute the cosmic 21-cm signal using the semi-numeric simulation \cmfst{}\footnote{https://github.com/andreimesinger/21cmFAST}\citep{Mesinger:2007p122, Mesinger:2011p1123}.  In this section, we provide a summary of the main recipes used to compute Equation~\ref{eq:21cmTb}, referring the reader to the aforementioned papers for more details.

\cmfst{} generates an evolved IGM density field (and associated velocity field) by perturbing an initially high resolution linear density field using second order LPT
\citep[e.g][]{Scoccimarro:1998p7939}
before smoothing it onto a lower resolution grid. Using an excursion-set approach \citep{Furlanetto:2004p123}, \cmfst{} then estimates the ionisation field from the 
evolved IGM density field by comparing the time-integrated number of ionising photons (produced by both UV and X-ray\footnote{Estimates based on local star-forming galaxies suggest that while X-rays only contribute to ionisations at the percent level, they dominate the IGM heating prior to the EoR, during the so-called EoH \citep[e.g.][]{Ricotti:2004p7145,Dijkstra:2012p7165,McQuinn:2012p3773,Mesinger:2013p1835}.}
sources) to the number of neutral atoms 
within regions of decreasing radius, $R$. 
This range starts from a maximum photon horizon, $R_{\rm mfp}$, down to the individual pixel resolution of a single voxel, $R_{\rm cell}$. Fully ionised 
voxels in the simulation volume are tagged when,
\begin{eqnarray} \label{eq:ioncrit}
\zeta f_{\rm coll}(\boldsymbol{x},z,R,\bar{M}_{\rm min}) \geq 1,
\end{eqnarray}
where $f_{\rm coll}(\boldsymbol{x},z,R,\bar{M}_{\rm min})$ is the fraction of collapsed matter residing within haloes more massive than $\bar{M}_{\rm min}$ \citep{Press:1974p2031,Bond:1991p111,Lacey:1993p115,Sheth:1999p2053} and $\zeta$ is an ionising efficiency describing the conversion of mass into ionising photons.

Equation (\ref{eq:ioncrit}) assumes a constant ionising luminosity to halo mass for the first galaxies, down to a threshold mass below which star-formation is inefficient (due to feedback and/or cooling).  This threshold mass, $\bar{M}_{\rm min}$, can also be represented in terms of a (redshift-dependent) halo virial temperature 
$T^{\rm min}_{\rm vir}$, \citep[e.g.][]{Barkana:2001p1634}
\begin{eqnarray}
M^{\rm min}_{\rm vir} &=& 10^{8} h^{-1} \left(\frac{\mu}{0.6}\right)^{-3/2}\left(\frac{\Omega_{\rm m}}{\Omega^{z}_{\rm m}}
\frac{\Delta_{\rm c}}{18\pi^{2}}\right)^{-1/2} \nonumber \\
& & \times \left(\frac{T^{\rm min}_{\rm vir}}{1.98\times10^{4}~{\rm K}}\right)^{3/2}\left(\frac{1+z}{10}\right)^{-3/2}M_{\sun},
\end{eqnarray}
where $\mu$ is the mean molecular weight, $\Omega^{z}_{\rm m} = \Omega_{\rm m}(1+z)^{3}/[\Omega_{\rm m}(1+z)^{3} + \Omega_{\Lambda}]$, 
and $\Delta_{c} = 18\pi^{2} + 82d - 39d^{2}$ where $d = \Omega^{z}_{\rm m}-1$.

As the 21cm signal is observed in contrast relative to the CMB, the IGM spin temperature, $T_{\rm S}$ (see Equation~\ref{eq:21cmTb}),
must be decoupled from the CMB temperature. This is achieved 
though either collisional coupling (effective mostly only during the Dark Ages) or through Wouthuysen-Field (WF) coupling 
\citep{Wouthuysen:1952p4321,Field:1958p1} to the \lya{} background generated from the first generation of stars.

\cmfst{} computes the IGM spin temperature in each simulation voxel by solving for the evolution in the ionisation, temperature and incident
\lya{} background\footnote{
Note, we do not vary the soft UV spectra of first sources responsible for setting this background as the WF coupling regime will likely
be beyond the observational capabilities of even the SKA and HERA. In a future publication, we will return to this.
}. These quantities depend on the angle-averaged specific intensity, $J(\boldsymbol{x}, E, z)$, (in erg s$^{-1}$ keV$^{-1}$ cm$^{-2}$ sr$^{-1}$),
which is computed by integrating the comoving X-ray specific emissivity, $\epsilon_{\rm X}(\boldsymbol{x}, E_e, z')$ back along the light-cone:
\begin{equation} \label{eq:Jave}
J(\boldsymbol{x}, E, z) = \frac{(1+z)^3}{4\pi} \int_{z}^{\infty} dz' \frac{c dt}{dz'} \epsilon_{\rm X}  e^{-\tau}.
\end{equation}
Here, the $e^{-\tau}$ term accounts for the probability of a photon emitted at an earlier time, $z'$, surviving until $z$ due to 
IGM attenuation (see Eq.\ 16 of \citealt{Mesinger:2011p1123}). The co-moving specific emissivity is, 
\begin{equation} \label{eq:emissivity}
\epsilon_{\rm X}(\boldsymbol{x}, E_{\rm e}, z') = \frac{L_{\rm X}}{\rm SFR} \left[ \rho_{\rm crit, 0} \Omega_b f_{\ast} (1+\delta_{\rm nl})\frac{d f_{\rm coll}(z')}{dt} \right],
\end{equation}
and is evaluated in the emitted frame, $E_{\rm e} = E(1 + z')/(1 + z)$. The quantity in square brackets is the star-formation rate (SFR) density back 
along the light-cone, with $\rho_{\rm crit, 0}$ being the current critical density and $f_\ast$ the fraction of galactic baryons converted into stars. 

The normalisation, $L_{\rm X}/{\rm SFR}$, is the specific X-ray luminosity per unit star formation escaping the host galaxies, $L_{\rm X}/{\rm SFR}$ 
(erg s$^{-1}$ keV$^{-1}$ $M^{-1}_{\odot}$ yr). We assume that the specific luminosity is a power-law in photon energy, $L_{\rm X} \propto E^{- \alpha_X}$, 
with photons below the threshold energy, $E_0$, being absorbed inside the host galaxy. The efficiency of the X-ray sources is expressed as an
integrated soft-band ($<2$~keV) luminosity per SFR (in erg s$^{-1}$ $M^{-1}_{\odot}$ yr):
\begin{equation} \label{eq:normL}
  L_{{\rm X}<2\,{\rm keV}}/{\rm SFR} = \int^{2\,{\rm keV}}_{E_{0}} dE_e ~ L_{\rm X}/{\rm SFR} ~,
\end{equation}
chosen to be more directly comparable with X-ray observations of local star-forming galaxies. This limit equally has a physical explanation, 
with harder photons having mean free paths longer than the Hubble length and hence  not contributing
to IGM heating during the EoR/CD \citep[e.g.][]{McQuinn:2012p3773,Das:2017p7170}.

\subsection{Redshift-space distortions}

The observed 21cm signal is impacted by line-of-sight peculiar velocities in two ways: (i) local peculiar velocity gradients distort the 
amplitude of the brightness temperature and (ii) observations are performed in {\it redshift} rather than {\it real} space, producing a distorted
map of the true cosmological signal. This latter effect shifts the {\it real} space co-moving signal to a {\it redshift} space signal 
along the line-of-sight through,
\begin{eqnarray} \label{eq:redspace}
\boldsymbol{s} = \boldsymbol{x} + \frac{(1+z)}{H(z)}v_{\parallel}(\boldsymbol{x}),
\end{eqnarray}
where $\boldsymbol{s}$ and $\boldsymbol{x}$ denotes {\it redshift} and {\it real} space signals, respectively. For an in-depth discussion
on RSDs, we refer the reader to \citet{Mao:2012p7838}. 
Previously, \cmfst{} accounted for
the first effect (e.g. the peculiar velocity gradient term in Equation~\ref{eq:21cmTb}), but not the second.
In order to mimic the observational signal as closely as possible over a wide range of models, here we additionally account for (ii).

Starting from the full expression of $\delta T_{\rm b}(\nu)$ (instead of the optically-thin approximation of Equation~\ref{eq:21cmTb}):
\begin{eqnarray}
\label{eq:21cmTb_Full}
\delta T_{\rm b}(\nu) = \frac{T_{\rm S} - T_{\rm CMB}}{1+z}\left( 1 - e^{-\tau_{\nu}}\right),
\end{eqnarray}
we implement redshift-space distortions following the simplified approach of \citet{Jensen:2013p1389}. This approach
closely resembles the recommended algorithm for semi-numerical simulations introduced by \citet{Mao:2012p7838}. We split each simulation voxel
into $n$ equally spaced sub-voxels along the line-of-sight, assigning to each sub-voxel a brightness temperature 
of $\delta T_{\rm b}(\boldsymbol{x})$/$n$. We then interpolate the velocity field onto the same sub-voxels, and perturb these according to 
Equation~\ref{eq:redspace}. Finally, we re-grid the sub-voxels back to the original resolution of the simulation.
We explore several choices of $n$, from 10, 20, 50 and 100, finding converged results for $n\gtrsim20$ both
in `by-eye' inspection of differential 21cm images (subtracting by the $n=100$ image) and statistics such as the 21cm PS.  Finally, we stitch together a light-cone by interpolating (linearly in cosmic time) between different snapshots of the redshift space distorted cubes (e,g, \citealt{Datta:2012p7679,Datta:2014p4990,LaPlante:2014p7651,Ghara:2015p7650,Mondal:2017p8569}).

\section{Mock Observations} \label{sec:obs}

\subsection{Fiducial cosmic signals} \label{sec:mock_models}

\begin{figure*} 
	\begin{center}
		\includegraphics[trim = 0cm 0.5cm 0cm 0.5cm, scale = 0.58]{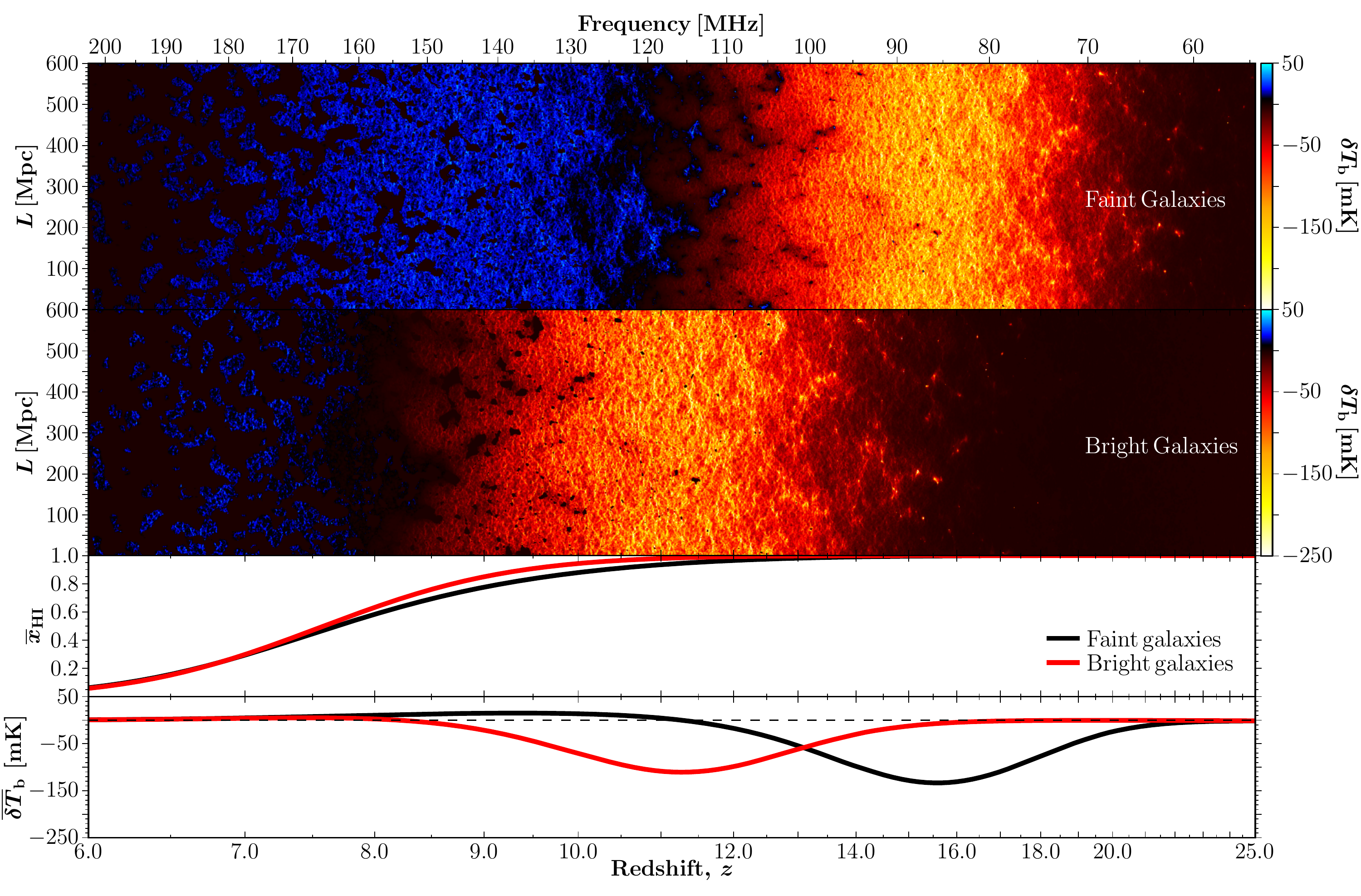}
	\end{center}
\caption[]
{A slice of the simulated 21cm brightness temperature light-cone for the \textsc{faint galaxies} (top panel) and the \textsc{bright galaxies} (second panel). These,
light-cones are generated from 400$^3$ co-eval cubes of length 600 Mpc.
The third and fourth panels correspond to the global neutral fraction ($\bar{x}_{\hi{}}$) and the globally averaged 21cm brightness temperature. 
See section~\ref{sec:mock_models} for more details about the two astrophysical models.}
 \label{fig:MockObs}
\end{figure*}

Following \citet{Greig:2017p8496}, we consider the same two mock observations, namely the \fg{} and \bg{} models. Primarily, these
two models differ owing to their choice of $T^{\rm min}_{\rm vir}$, which controls the timing of the cosmic epochs, and the bias of the sources
responsible. The two choices bracket the plausible scatter in the cosmic SFR density inferred from extrapolating the faint end of the high-$z$ 
luminosity functions (c.f. \citealt{Bouwens:2015p7832}, Fig. 5 in \citealt{Das:2017p7170}).
The ionising efficiency, $\zeta$, was then varied to ensure each model matched 
the electron scattering optical depth, $\tau_{e}$, 
constraints from Planck ($\tau_{\rm e} = 0.058 \pm 0.012$; \citealt{PlanckCollaboration:2016p7627}). Specifically,
\begin{enumerate}
\item[\fg{}:] $T^{\rm min}_{\rm vir} = 5\times10^4$~K, $\zeta$ = 30, $\tau_{\rm e} = 0.059$,
\item[\bg{}:] $T^{\rm min}_{\rm vir} = 3\times10^5$~K, $\zeta$ = 200, $\tau_{\rm e} = 0.056$,
\end{enumerate}
with the full astrophysical parameter set provided in the first row of Tables~\ref{tab:FaintGalaxies} and~\ref{tab:BrightGalaxies}.

In Figure~\ref{fig:MockObs}, the 
upper two panels correspond to a slice through the 3D light-cone, for the \fg{} and \bg{} models respectively. In the third and final panels, we provide the
global averaged evolution in the fraction of neutral hydrogen ($\bar{x}_{\hi{}}$) and the global average 21cm brightness temperature ($\bar{\delta T_{\rm b}}$).
This mock light-cone observation was generated from co-eval cubes with a volume of 600$^{3}$ Mpc$^{3}$, on a 400$^3$ grid smoothed down 
from the high-resolution initial conditions generated on a 2400$^3$ grid. Importantly, these are generated from initial conditions using a different random seed
than used for the MCMC sampling.

Since the ionising efficiency is adjusted to match the observed CMB optical depth, the major difference between the two models is the timing of the EoH.  In the \textsc{bright galaxies} model, the EoH occurs late, overlapping significantly with the  EoR.  This is driven by the relatively late but rapid emergence of the rare, biased sources in this model.  We expect this rapid evolution of the \textsc{bright galaxies} model to result in the largest biases when performing parameter recovery from co-eval cubes instead of directly on the light-cone.

\subsection{Telescope noise}

As we are investigating the impact of the light-cone effect in the context of astrophysical parameter recovery, we must model the expected 
noise from 21 cm experiments. Again, we focus solely on the 21cm PS, and use the python module 
\sense{}\footnote{https://github.com/jpober/21cmSense}\citep{Pober:2013p41,Pober:2014p35} to estimate the 21cm PS sensitivity curves. We
summarise the general method below.

The thermal noise PS is generated from the gridded $uv$-visibilities according to the following \citep[e.g.][]{Morales:2005p1474,McQuinn:2006p109,Pober:2014p35},
\begin{eqnarray} \label{eq:NoisePS}
\Delta^{2}_{\rm N}(k) \approx X^{2}Y\frac{k^{3}}{2\pi^{2}}\frac{\Omega^{\prime}}{2t}T^{2}_{\rm sys},
\end{eqnarray} 
where $X^{2}Y$ is a cosmological conversion factor between observing bandwidth, frequency and co-moving distance, $\Omega^{\prime}$ is a 
beam-dependent factor derived in \citet{Parsons:2014p781}, $t$ is the total time spent by all baselines within a particular $k$-mode and 
$T_{\rm sys}$ is the system temperature, the sum of the receiver temperature, $T_{\rm rec}$, and the sky temperature $T_{\rm sky}$. 
We model $T_{\rm sky}$ using the frequency dependent scaling $T_{\rm sky} = 60\left(\frac{\nu}{300~{\rm MHz}}\right)^{-2.55}~{\rm K}$ \citep{Thompson2007}.

The sample variance is then estimated from the cosmological 21 cm PS (mock observation, $\Delta^{2}_{21}(k)$) and 
combined with the thermal noise using an inverse-weighted summation over all the 
individual modes \citep{Pober:2013p41}. The total noise power, $\delta\Delta^{2}_{\rm T+S}(k)$, at a given Fourier mode, $k$, is then,
\begin{eqnarray} \label{eq:T+S}
\delta\Delta^{2}_{\rm T+S}(k) = \left(\sum_{i}\frac{1}{(\Delta^{2}_{{\rm N},i}(k) + \Delta^{2}_{21}(k))^{2}}\right)^{-\frac{1}{2}}.
\end{eqnarray}
Here we assume Gaussian errors for the cosmic-variance term, which is a good approximation on large-scales.

Throughout, we use the ``moderate'' foreground removal strategy of \citet{Pober:2014p35}. This confines the 21cm signal to the 
pristine 21 cm window (i.e. avoiding the contaminated foreground ``wedge''), placing the wedge at 
$\Delta k_{\parallel} = 0.1 \,h$~Mpc$^{-1}$ beyond the horizon limit \citep{Pober:2014p35}. This strategy also assumes 
coherent summation over all redundant baselines, reducing the resultant thermal noise \citep{Parsons:2012p95}.

We must be careful at which redshift we estimate the instrumental sensitivity. Ideally, we would populate a corresponding 3D noise cuboid with
each 2D slice along the line-of-sight generated from the noise PS at each different observing frequency.  Then we could extract the noise in the same manner as we do for the signal, preserving the evolution of the noise along the light-cone.
However, for this proof-of-concept work, we simply use co-eval cubes of noise generated at the central redshift of the light-cone slices we use for the 21-cm PS (see below).
We do not expect this simplification to significantly impact our results, as
the thermal noise does not vary as quickly as the 21cm signal along the line-of-sight.

Finally, we use HERA for our archetypal telescope when computing sensitivities.  HERA and the SKA are the only planned interferometers with sufficient sensitivity to measure power spectra over the entire redshift range shown in Fig. 1. We expect the results for the SKA to be quite comparable \citep[e.g.][]{Greig:2017p8496}, although we do not compute them explicitly so as to not complicate the presentation,

We model HERA following the earlier design outlined in \citet{Beardsley:2014p1529} with a core design consisting of 331 dishes\footnote{
Note, the final HERA design will consist of 350 dishes, 320 in the core with 30 outrigger stations \citep{DeBoer:2017p6740}. The 
sensitivity difference on the 21cm PS between a 320 and 331 core layout should be negligible. }
Each parabolic dish is 14m in diameter, densely packed in a hexagonal configuration to maximise the total number of 
redundant baselines \citep{Parsons:2012p95}. Operating in a drift-scanning mode, we assume a total 1080~hr observation, 
spread across 180 nights at 6 hours per night and a total system temperature of $T_{\rm sys} = 100 + T_{\rm sky}~{\rm K}$. 
The total observing bandwidth of HERA covers 50 - 250 MHz.

%%%%%%%%%%%%%%%%%%%%%%%%%%%%%%%%%%%%%%%%%%%%%%%%%%%%%%%%%%%%%%%%%%%%%%%%%%%%%%%%%%%%%%%%%%%%%%%%%%%%%%%%%%%%%%%%%%%%%%%%%%%%%%%%
%%%%%%%%%%%%%%%%%%%%%%%%%%%%%%%%%%%  PARAMETER EXTRACTION  %%%%%%%%%%%%%%%%%%%%%%%%%%%%%%%%%%%%%%%%%%%%%%%%%%%%%%%%%%%%%%%%%%%
%%%%%%%%%%%%%%%%%%%%%%%%%%%%%%%%%%%%%%%%%%%%%%%%%%%%%%%%%%%%%%%%%%%%%%%%%%%%%%%%%%%%%%%%%%%%%%%%%%%%%%%%%%%%%%%%%%%%%%%%%%%%%%%%
\section{Parameter estimation on the light-cone} \label{sec:21cmmc}

In the previous sections, we discussed the mock observations, and how a cosmic signal is generated for a given set of astrophysical parameters with \cmfst{}.  Here we summarise how we perform the sampling of astrophysical parameters and statistical comparison to the mock observations, which results in our parameter constraints.

For our Bayesian sampler, \cmmc{}, which is a massively parallel MCMC sampler of an optimised version of \cmfst{}.
Developed from a modified version of the \CH{} python module \citep{Akeret:2012p842}, it uses the \emcee{} python module \citep{ForemanMackey:2013p823}, an affine
invariant ensemble sampler \citep{Goodman:2010p843}. \cmmc{} performs a new, independent 3D realisation of the 21 cm signal at each proposed MCMC step, and uses 
a likelihood statistic (fiducially the 21cm PS) to compare the sampled model against a mock observation. 

\subsection{Astrophysical parameter set} \label{sec:param}

To facilitate comparison, we sample the same astrophysical parameter set used in \citet{Greig:2017p8496}.  We briefly summarise these parameters here, referring the reader to \citet{Greig:2017p8496} for further details.\footnote{We also provide animations highlighting the impact of each astrophysical parameter on the observed 21cm light-cone
at http://homepage.sns.it/mesinger/21CMMC.html}.

\begin{itemize}
  \item $\zeta$ - {\it the UV ionising efficiency of high-$z$ galaxies}. The ionising efficiency (c.f. Equation~\ref{eq:ioncrit}) can be expressed as 
 \begin{eqnarray} \label{eq:Zeta}
\zeta = 30\left(\frac{f_{\rm esc}}{0.12}\right)\left(\frac{f_{\ast}}{0.05}\right) \left(\frac{N_{\gamma/b}}{4000}\right)\left(\frac{1.5}{1+n_{\rm rec}}\right)
 \end{eqnarray}
 where, $f_{\rm esc}$ is the fraction of ionising photons escaping into the IGM, $f_{\ast}$ is the fraction of galactic gas in stars, $N_{\gamma/b}$ is the number of ionising photons produced per baryon in stars and $n_{\rm rec}$ is the typical number of times a hydrogen atom recombines.  $\zeta$ controls the timing of reionisation, with larger values shifting reionisation to earlier times while also shortening its duration. Within \cmmc{} we adopt a flat prior over the fiducial range of $\zeta\in[10,250]$.
\item $R_{\rm mfp}$: {\it maximum ionising photon horizon}.
  Photon sinks (e.g.\ Lyman limit systems) and the associated recombinations limit the extent to which photons can propagate through the {\it ionised} IGM.  As is common in the literature, we approximate this effect with a maximum horizon through the ionised IGM, $R_{\rm mfp}$ (c.f. the discussion surrounding eq. \ref{eq:ioncrit}).  We note that $R_{\rm mfp}$ has a minimal impact on reionisation, potentially slowing down the final stages when the typical \hii{} region scales grow to exceed $R_{\rm mfp}$ \citep[e.g.][]{Furlanetto:2005p4326,Alvarez:2012p1930}.
  In this work, we adopt a flat prior over $R_{\rm mfp}\in[5, 25]$~cMpc
consistent with the \citet{Sobacchi:2014p1157} sub-grid recombination model. 
  
\item $T^{\rm min}_{\rm vir}$: {\it minimum virial temperature of star-forming haloes}.
  Galaxies hosted in haloes with virial temperatures above $T^{\rm min}_{\rm vir}$ are assumed to have a constant ionising luminosity to halo mass, while galaxies below this threshold are assumed to have negligible star-formation (due to negative feedback and/or difficulty in cooling).  Star-forming galaxies are assumed to be the sources of all of the radiation fields which govern the cosmic 21-cm signal.  Therefore, unlike the EoR parameters above, $T^{\rm min}_{\rm vir}$ impacts the timing and source bias of {\it all} cosmic epochs. Higher values result in later, more rapid cosmic evolutions.
We take a flat prior across $T^{\rm min}_{\rm vir}\in[10^{4},10^{6}]$~K within this work.\footnote{Our upper limit is consistent with the host halo masses inferred through the abundance matching of observed
$z\sim6-8$ Lyman break galaxies (e.g. \citealt{Kuhlen:2012p1506, BaroneNugent:2014p4324}).
The lower limit, is motivated by the minimum temperature for efficient atomic line cooling.
Though, $T^{\rm min}_{\rm vir}$ 
can be as low as $\approx10^{2}$~K in the presence of radiative cooling 
\citep{Haiman:1996p2144,Tegmark:1997p7180,Abel:2002p2149,Bromm:2002p2153}, star-formation 
within these haloes is likely inefficient (a few stars per halo; e.g.~\citealt{Kimm:2017p7875}) and can quickly 
($z>20$) be suppressed by Lyman-Werner or other feedback processes well before the EoR 
\citep{Haiman:2000p2155,Ricotti:2001p2160,Haiman:2006p2169,Mesinger:2006p2171,Holzbauer:2012p2890,Fialkov:2013p2903}.}

\item $L_{\rm X\,<\,2\,keV}/{\rm SFR}$: {\it the integrated soft-band X-ray luminosity escaping galaxies.}
The efficiency of X-rays at heating the IGM preceding the EoR is driven by the total integrated soft-band ($<$~2~keV) luminosity per 
SFR (Equation~\ref{eq:normL}) escaping the host galaxies. $L_{\rm X\,<\,2\,keV}/{\rm SFR}$ regulates the timing of the EoH (analogous to the role of $\zeta$ for the EoR).
We adopt a flat prior over the range ${\rm log}_{10}(L_{\rm X\,<\,2\,keV}/{\rm SFR})\in[38,42]$,
consistent with predictions from population synthesis models applied at high-redshifts \citep{Fragos:2013p6529} and conservatively broader
than individual objects observed locally \citep{Mineo:2012p6282} or in stacked observations \citep{Lehmer:2016p7810}. 

\item $E_{0}$: {\it X-ray energy threshold for self-absorption by the host galaxies.}
The interstellar medium (ISM) of high-$z$ galaxies can absorb soft X-ray photons, attenuating the emergent X-ray profile, and limiting IGM heating.
\citet{Das:2017p7170} find that the emergent X-ray spectrum from simulated high-$z$ galaxies can be well fit by a step-function attenuation below $E_{0}\sim0.5$ keV (corresponding to a metal free ISM with a \hi{} column density of ${\rm log_{10}}(N_{\rm \hi{}}/{\rm cm^{2}}) = 21.5$).  At a fixed soft-band luminosity, increasing $E_0$ hardens the emerging spectrum.  Since the mean free path is a strong function of photon energy, the resulting EoH is more uniform.
Here we take a flat prior over $E_{0}\in[0.1,1.5]$~keV which corresponds approximately to an optical depth of unity through a metal free ISM with ${\rm log_{10}}(N_{\rm \hi{}}/{\rm cm^{2}})\in[19.3,23.0]$.
  
\item $\alpha_{\rm X}$: {\it X-ray spectral energy index.}
This depends on the dominant physical 
process producing the X-ray photons. 
Increasing/decreasing $\alpha_{\rm X}$ results in more numerous soft/hard X-ray
photons (when the soft-band luminosity is fixed), resulting in similar qualitative behaviour to that for decreasing/increasing $E_{0}$.
We adopt a flat prior of $\alpha_{\rm X}\in[-1.0,3.0]$, which encompasses a wide range of plausible 
X-ray spectral energy distributions that describe high-$z$ galaxies (e.g. HMXBs, host ISM, mini-quasars, supernovae remnants etc.; 
see for example \citealt{McQuinn:2012p3773,Pacucci:2014p4323}).

\end{itemize}

\subsection{Compressing the 21cm light-cone into summary statistics} \label{sec:lc}

Currently, there is no general consensus on the optimal method for 
analysing light-cone  data. \citet{Datta:2012p7679}, point out that 
the light-cone and co-eval 21cm signal converge for
decreasing bandwidth. However, this reduces the amount of large-scale information, which typically is the most sensitive to the 
underlying astrophysics \citep[see e.g. Figure 2 of][]{Greig:2015p3675}.

Alternately, \citet{Trott:2016p7921} propose the use of a wavelet transform along the line-of-sight.  Compared to carving up the light-cone with sharp boundaries (i.e. a box-car filter), the wavelet transform preserves some of the large-scale line of sight information underneath a damped tail. In their toy examples, they find improved statistical performance compared to a box-car filter.

In this proof-of-concept work, we take the simple approach of splitting the light-cones into equal co-moving distance boxes, and generate a 21cm PS
from each.  We use the same co-moving scale for both the mock observation and the MCMC samples.  In future, we will return to this issue by exploring the impact of the more optimal wavelet transform method.

\begin{table*}
\begin{tabular}{@{}lcccccccccc}
\hline
\textsc{faint galaxies} & & & & Parameter & &  \\
Model & $\zeta$ & $R_{\rm mfp}$& log$_{10}$$(T^{\rm min}_{\rm vir})$ & log$_{10}$($L_{\rm X\,<\,2\,keV}/{\rm SFR}$) & $E_{0}$ & $\alpha_{X}$ & $\chi^{2}_{\rm min}$ \\
HERA 331 &  & [Mpc] & [K] & [erg s$^{-1}$ $M^{-1}_{\odot}$~yr] & [keV] & \\
\hline
\vspace{0.8mm}
Fiducial values & 30.0 & 15.0 & 4.70 & 40.0 & 0.5 & 1.0 & - \\
\hline
\vspace{0.8mm}
Light-cone & $32.20\substack{+2.68 \\ -2.08}$ & $14.86\substack{+0.78 \\ -0.86}$ & $4.74\substack{+0.04 \\ -0.03}$ & $40.06\substack{+0.05 \\ -0.06}$ & $0.50\substack{+0.09 \\ -0.14}$ & $0.63\substack{+1.17 \\ -0.92}$ & 8.60\\
\hline
\vspace{0.8mm}
Co-eval cubes & $31.28\substack{+2.32 \\ -1.87}$ & $14.77\substack{+0.81 \\ -0.80}$ & $4.72\substack{+0.04 \\ -0.03}$ & $40.03\substack{+0.05 \\ -0.05}$ & $0.42\substack{+0.04 \\ -0.05}$ & $1.15\substack{+0.52 \\ -0.44}$ & 18.14\\
\hline
\end{tabular}
\caption{Summary of the \fg{} astrophysical model (top row), and the median recovered values (and associated 16th and 84th percentile errors) assuming the 
full light-cone signal (middle row) or the co-eval approximation (bottom row). For each, we have assumed a total 1000hr integration time with HERA. Note, we would anticipate
similar behaviour when considering the SKA, owing to the similar performance between HERA and the SKA in \citet{Greig:2017p8496}.
}
\label{tab:FaintGalaxies}
\end{table*}

%%%%%%%%%%%%%%%%%%%%%%%%%%%%%%%%%%%%%%%%%%%%%%%%%%%%%%%%%%%%%%%%%%%%%%%%%%%%%%%%%%%%%%%%%%%%%%%%%%%%%%%%%%%%%%%%%%%%%%%%%%%%%%%%
%%%%%%%%%%%%%%%%%%%%%%%%%%%%%%%%%%%  RESULTS  %%%%%%%%%%%%%%%%%%%%%%%%%%%%%%%%%%%%%%%%%%%%%%%%%%%%%%%%%%%%%%%%%%%
%%%%%%%%%%%%%%%%%%%%%%%%%%%%%%%%%%%%%%%%%%%%%%%%%%%%%%%%%%%%%%%%%%%%%%%%%%%%%%%%%%%%%%%%%%%%%%%%%%%%%%%%%%%%%%%%%%%%%%%%%%%%%%%%
%\section{Results}

\section{Parameter forecasts from the light-cone} \label{sec:Forecasts}

\subsection{\cmmc{} setup}

The light-cones sampled within the MCMC are generated from 300$^{3}$ Mpc$^{3}$ co-eval cubes, on a 200$^3$ grid, 
smoothed down from a high-resolution 1200$^{3}$ grid. These, retain the same spatial resolution ($\sim1.5$~Mpc per voxel) as
the mock observations. Our lightcones extend from
$z=6$ ($\sim200$~MHz) to $z=26.8$ ($\sim50$~MHz).  This results in a total of 10 independent ``chunks'' which are used to construct power spectra.

In previous studies, we
used $k=0.15$~Mpc$^{-1}$ as the lower limit for the fitting range of the 21cm PS, assuming larger scales are lost to foreground contamination.  However, in this work we are more optimistic about foreground mitigation, extending our PS range down to 
$k=0.1$~Mpc$^{-1}$, while keeping the same upper limit of $k=1.0$~Mpc$^{-1}$.
The inclusion of more large-scale modes, which are the ones which evolve most strongly along the line-of-sight, highlights the impact of the light-cone effect on the astrophysical parameter recovery. Furthermore, we also decrease the modeling uncertainty from 20\% to 10\% here\footnote{
This modelling uncertainty accounts for inaccuracies in semi-numerical approaches such as \cmfst{}. We assume a fixed, uncorrelated 
percentage error, whose amplitude is roughly motivated by comparisons against the radiative transfer simulations in \citet{Zahn:2011p1171}.
}. 
These two optimistic choices enhance the importance of the light-cone effect; therefore our results below can be taken as upper limits on the parameter bias from ignoring the light-cone.

Generating the full 3D light-cone within \cmmc{} is more computationally expensive than the previous version. This is because
the ionisation field must be generated for all co-eval boxes to construct the light-cone, rather than only a handful of select co-eval redshifts.
In the end, for the same box size and resolution, we find that \cmmc{} with a light-cone is $\sim2$ times slower. Our \cmmc{}
setup was: 324 walkers with 250 iterations each, resulting in $\sim8\times 10^{4}$ samplings of the 3D light-cone. Performed
using 162 cores on a shared memory cluster, such a setup takes $\sim18$ days.
This time can be reduced by a factor of $\sim$2--4 by decreasing the resolution from 200$^3$ to 128$^3$. Although convergence tests suggest 128$^3$ is sufficient for accurate parameter recovery, here we retain 200$^3$ to match our previous studies.

\subsection{Faint Galaxies} \label{sec:fg}

\begin{figure*} 
	\begin{center}
		\includegraphics[trim = 0.1cm 0.6cm 0cm 0.3cm, scale = 0.88]{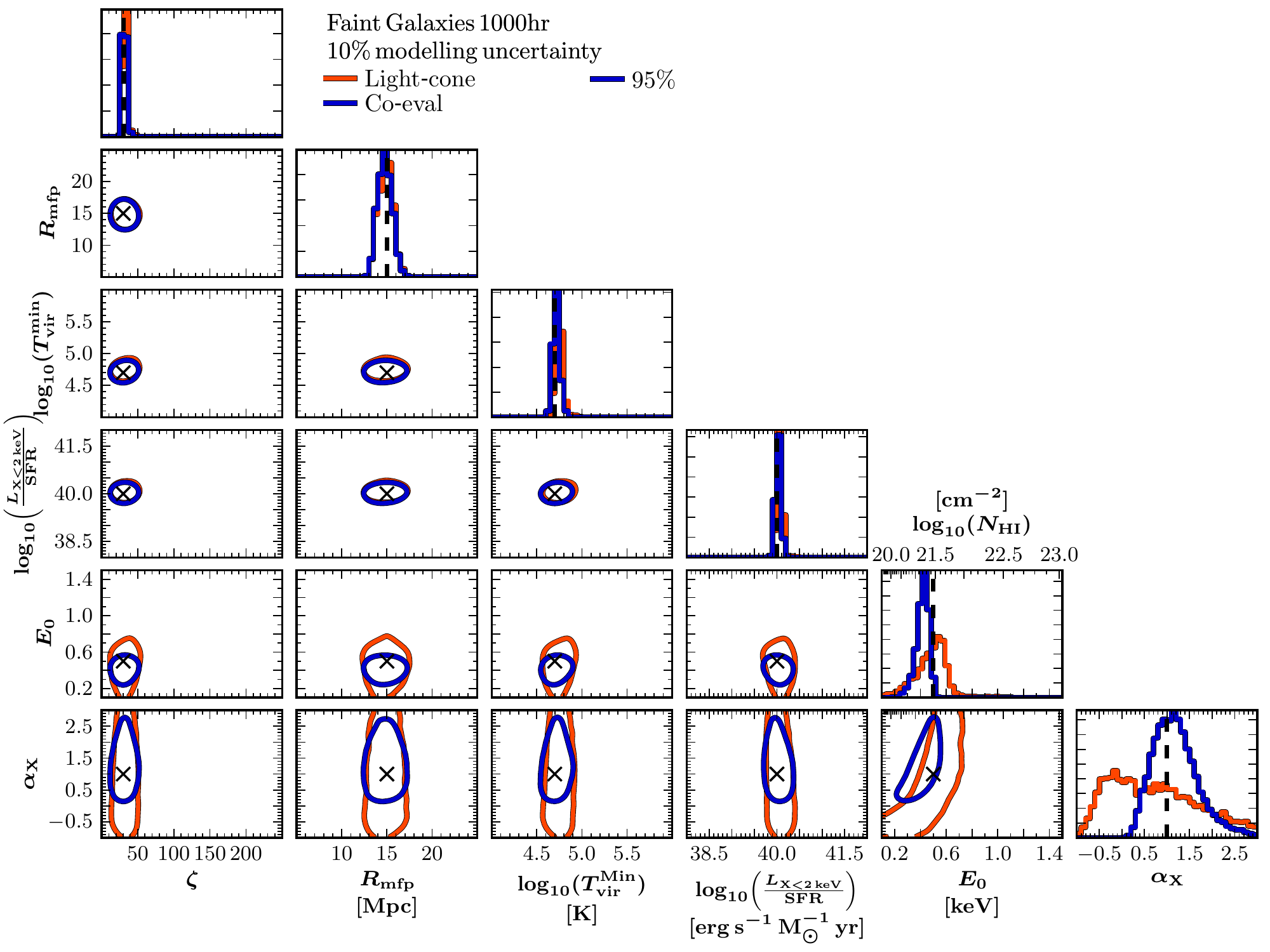}
	\end{center}
\caption[]
{Recovered 2D marginalised joint posterior distributions for our \textsc{faint galaxies} mock light-cone observation, for an assumed 1000 hr on sky observation with HERA. The different colours denote fitting the mock observation with the true light-cone signature (red) or assuming co-eval evolution of the 21cm signal only (blue). Black crosses (and vertical dashed lines) correspond to the input model parameters, defined to be ($\zeta$, $R_{\rm mfp}$, log$_{10}$$(T^{\rm min}_{\rm vir})$, log$_{10}$($L_{\rm X\,<\,2\,keV}/{\rm SFR}$), $E_{0}$, $\alpha_{X}$) = (30, 15, 4.7, 40.0, 0.5, 1.0). 
}
 \label{fig:LikelihoodFaint}
\end{figure*}

In Figure~\ref{fig:LikelihoodFaint}, we present the astrophysical parameter constraints for the \fg{} model assuming a 1000hr observation with HERA,
using the 21cm PS from the observed light-cone (red). Diagonal panels correspond to the normalised 1D marginalised probability distribution functions (PDFs)
for each of the six astrophysical parameters (see Section~\ref{sec:param}). In the lower left half of the figure, we provide the 95 per cent 2D marginalised joint likelihood contours, with crosses denoting the fiducial parameters of the \fg{} model. In Table~\ref{tab:FaintGalaxies} we provide the 
median and associated 16th and 84th percentiles for each of the six astrophysical parameters (middle row), with the fiducial \fg{} parameter set 
in the top row. We also provide the resultant $\chi^{2}_{\rm min}$ value for the \fg{} model in the final column.

In order to have a convenient figure-of-merit for the recovery, we approximate the marginalised 1D PDFs as normal distributions 
(which is a reasonable approximation at the 1$\sigma$ level). The 1$\sigma$ percentage errors are then: 
($\zeta$, $R_{\rm mfp}$, log$_{10}$$(T^{\rm min}_{\rm vir})$, log$_{10}$($L_{\rm X\,<\,2\,keV}/{\rm SFR}$), $E_{0}$, $\alpha_{X}$) = 
(7.4, 5.5, 0.7, 0.1, 23.0, 165.8) per cent respectively.
Consistent with \citet{Greig:2017p8496}, we are able to constrain most parameters to $\lsim10$ per cent level precision, aside from $\alpha_X$.  However, the light-cone does ``smear-out'' some of the large-scale modes during the EoH which can constrain the X-ray SED.  Thus, the  $E_{0}$--$\alpha_X$ degeneracy is stronger than found by \citet{Greig:2017p8496}.

\begin{table*}
\begin{tabular}{@{}lcccccccccc}
\hline
\textsc{bright galaxies} & & & & Parameter & &  \\
Model & $\zeta$ & $R_{\rm mfp}$& log$_{10}$$(T^{\rm min}_{\rm vir})$ & log$_{10}$($L_{\rm X\,<\,2\,keV}/{\rm SFR}$) & $E_{0}$ & $\alpha_{X}$ & $\chi^{2}_{\rm min}$ \\
HERA 331 &  & [Mpc] & [K] & [erg s$^{-1}$ $M^{-1}_{\odot}$~yr] & [keV] & \\
\hline
\vspace{0.8mm}
Fiducial values & 200.0 & 15.0 & 5.48 & 40.0 & 0.5 & 1.0 & - \\
\hline
\vspace{0.8mm}
Light-cone & $196.01\substack{+16.23 \\ -17.04}$ & $14.56\substack{+0.85 \\ -0.77}$ & $5.48\substack{+0.01 \\ -0.01}$ & $40.04\substack{+0.05 \\ -0.05}$ & $0.54\substack{+0.05 \\ -0.07}$ & $0.68\substack{+0.95 \\ -0.90}$  & 14.34\\
\hline
\vspace{0.8mm}
Co-eval cubes & $114.26\substack{+5.72 \\ -5.32}$ & $22.88\substack{+0.93 \\ -1.16}$ & $5.40\substack{+0.01 \\ -0.01}$ & $39.92\substack{+0.02 \\ -0.02}$ & $0.18\substack{+0.04 \\ -0.05}$ & $0.15\substack{+0.07 \\ -0.08}$  & 44.55\\
\hline
\end{tabular}
\caption{Same as Table~\ref{tab:FaintGalaxies} except now for the \textsc{bright galaxies} model which assumes ($\zeta$, $R_{\rm mfp}$, log$_{10}$$(T^{\rm min}_{\rm vir})$, log$_{10}$($L_{\rm X\,<\,2\,keV}/{\rm SFR}$), $E_{0}$, $\alpha_{X}$) = (200, 15, 5.48, 40.0, 0.5, 1.0).}
\label{tab:BrightGalaxies}
\end{table*}

\begin{figure*} 
	\begin{center}
		\includegraphics[trim = 0.1cm 0.6cm 0cm 0.5cm, scale = 0.88]{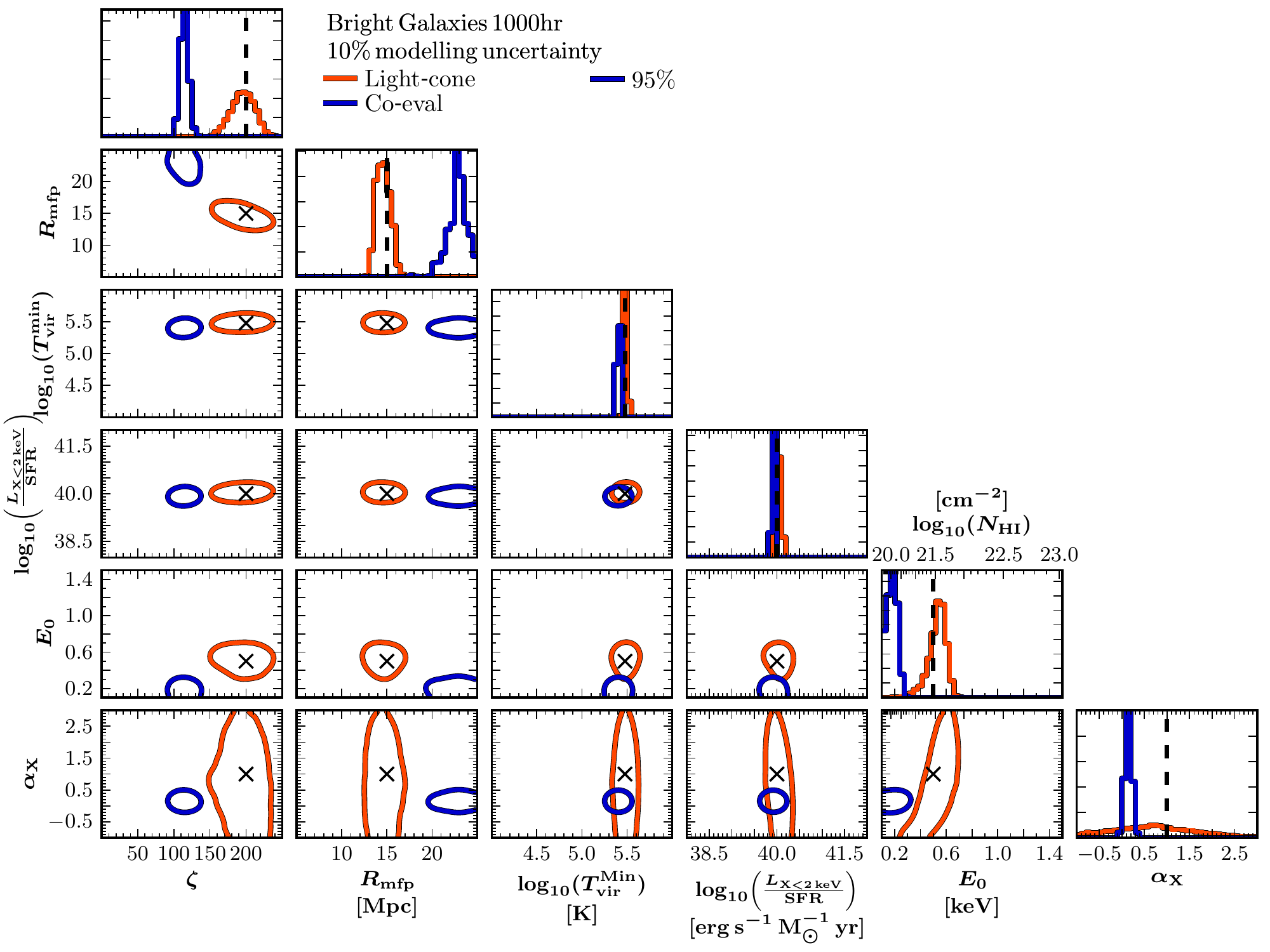}
	\end{center}
\caption[]
{The same as Figure~\ref{fig:LikelihoodFaint} except for the \textsc{bright galaxies} mock light-cone observation. Here, the fiducial input parameters correspond to ($\zeta$, $R_{\rm mfp}$, log$_{10}$$(T^{\rm min}_{\rm vir})$, log$_{10}$($L_{\rm X\,<\,2\,keV}/{\rm SFR}$), $E_{0}$, $\alpha_{X}$) = (200, 15, 5.48, 40.0, 0.5, 1.0).}
 \label{fig:LikelihoodBright}
\end{figure*}

\subsection{Bright Galaxies} \label{sec:bg}

In Figure~\ref{fig:LikelihoodBright}, we present the corresponding astrophysical parameter constraints for the \bg{} model using the full light-cone (red curves) assuming the same 
1000hr observation with HERA. In the middle row of Table~\ref{tab:BrightGalaxies} we provide the median and associated 16th and 84th percentiles for each astrophysical parameter.
As in the previous section, the approximate 1$\sigma$ percentage errors are: 
($\zeta$, $R_{\rm mfp}$, log$_{10}$$(T^{\rm min}_{\rm vir})$, log$_{10}$($L_{\rm X\,<\,2\,keV}/{\rm SFR}$), $E_{0}$, $\alpha_{X}$) = 
(8.5, 5.6, 0.2, 0.1, 11.0, 136.0) per cent respectively.

Four of the six astrophysical parameters between the \fg{} and \bg{} models are kept fixed 
(e.g. $R_{\rm mfp}$, log$_{10}$($L_{\rm X\,<\,2\,keV}/{\rm SFR}$), $E_{0}$, $\alpha_{X}$). Notably, these including the three X-ray 
parameters that describe the epoch of heating. While $L_{\rm X\,<\,2\,keV}/{\rm SFR}$ is constrained to equally high precision between both
models, we find that the constraints on $E_{0}$ are improved by a factor of two, with $\alpha_{X}$ also exhibiting some improvement.

In the \bg{} model, IGM heating by X-rays occurs at later times
owing to the rarer, more biases sources considered (Figure~\ref{fig:MockObs}). Since the thermal sensitivity of a 21cm interferometer scales 
inversely with observing frequency, the heating peak is shifted into a regime of lower thermal noise, increasing the sensitivity of the signal 
and thus the constraints on the X-ray parameters.

\section{Can ignoring the light-cone effect bias astrophysical parameter recovery?} \label{sec:bias}

In this section, we use \cmmc{} to quantify the parameter bias resulting from fitting our 
mock \fg{} and \bg{} light-cone observations using the common co-eval approximation in the MCMC.
We calculate the 21cm PS from the same number of co-eval cubes
as the number of equal co-moving distance cubes extracted from the light-cone (i.e. 10), and take the central redshift of each of the light-cone cubes
to be the redshift of each co-eval cube.  Thus the co-eval cubes are located at $z=6.4$, 7.2, 8.1, 9.2, 10.7, 12.2, 14.5, 16.8, 20.2 and 24.4.

\begin{figure*} 
	\begin{center}
		\includegraphics[trim = 0cm 0.5cm 0cm 0.5cm, scale = 0.9]{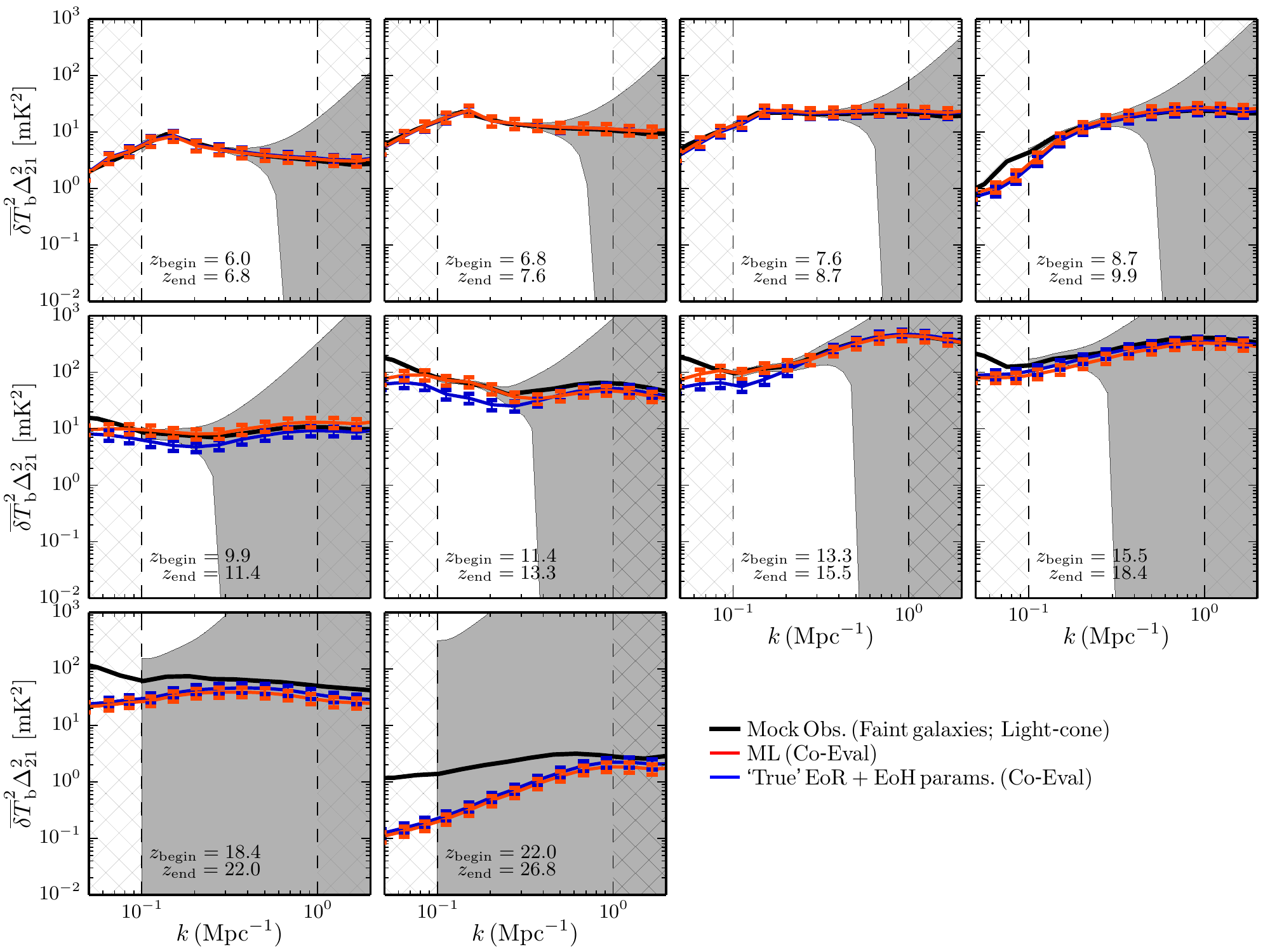}
	\end{center}
\caption[]
{The maximum likelihood (ML) 21cm PS obtained from assuming co-eval simulation curves (red curve) compared to the fiducial mock 21cm PS generated from the \textsc{faint galaxies} mock light-cone observation (black). For reference, the blue curves correspond to the ``true'' EoR and EoH model parameters, ($\zeta$, $R_{\rm mfp}$, log$_{10}$$(T^{\rm min}_{\rm vir})$, log$_{10}$($L_{\rm X\,<\,2\,keV}/{\rm SFR}$), $E_{0}$, $\alpha_{X}$) = (30, 15, 4.7, 40.0, 0.5, 1.0), but in the co-eval approximation. The grey shaded region corresponds to the 1$\sigma$ observational uncertainty for an assumed 1000hr observation with HERA, while the error bars denote our assumed 10 per cent modelling uncertainty on the 21 cm PS. Hatched regions denote $k$-modes outside of our nominal fitting range.}
 \label{fig:BiasFaint}
\end{figure*}

\subsection{Faint Galaxies}

In the bottom row of Table~\ref{tab:FaintGalaxies}, we provide the median and associated 16th and 84th percentiles for the astrophysical
parameters assuming the co-eval approximation, while the blue curves in Figure~\ref{fig:LikelihoodFaint}, 
correspond to the marginalised 1 and 2D posterior distributions. 
Immediately evident from Figure~\ref{fig:LikelihoodFaint} is that there are no biases on the recovered astrophysical parameters. Equally,
for the EoR parameters (i.e.\ $\zeta$, $R_{\rm mfp}$ and $T^{\rm min}_{\rm vir}$) the recovered precision remains unaffected by ignoring
the light-cone effect. However, for the IGM heating parameters, the widths of the
recovered 1D posteriors ($E_{0}$ and $\alpha_{X}$) have notably narrowed by ignoring the light-cone effect. In the co-eval approximation,
the recovered percentage errors on $E_{0}$ and $\alpha_{X}$ are 11.1 and 41.7 per cent, compared to 23.0 and 165.8 per cent with the light-cone.
This corresponds to an over-estimation of the true precision available with HERA by a factor of 2 and 4, respectively. Although ignoring
the light-cone effect did not bias the astrophysical parameters, they still lead to strongly overestimating the true precision to which these parameters can 
be recovered from 21cm experiments.

To analyse this further, in Figure~\ref{fig:BiasFaint}, we present the 21cm PS from the \fg{} mock observation 
for each of the equal co-moving light-cone boxes (black). We then compare the maximum likelihood (ML) 21cm PS obtained from \cmmc{} 
when adopting the co-eval approximation (red curve). Finally, the blue curves correspond to the 21cm PS from the ``true'' \fg{} astrophysical model,
assuming the co-eval approximation.

Along the top row, corresponding to $z=6-10$, we find the 21cm PS obtained from the co-eval approximation to match extremely well with the light-cone 21cm PS.
This effectively spans the entirety of the EoR (at $z=10$, $\bar{x}_{\hi{}} \sim 0.85$), explaining why the EoR parameter constraints are unaffected by the 
light-cone effect. The full redshift span across each of these boxes is $\Delta z \sim 0.8 - 1.2$, corresponding to an observing 
frequency of $\Delta \nu \sim 16 - 21$~MHz. However, given our sampled fitting region is $k = 0.1 - 1.0$~Mpc$^{-1}$ the effective redshift (frequency) span 
of these modes is further reduced, and consistent with the optimal bandwidths discussed in \citet{Datta:2014p4990}.

Beyond $z=10$, notable differences begin to occur. While the ML 21cm PS (red curves) are still capable of matching the light-cone 21cm PS extremely well, 
the 21cm PS from the \fg{} model parameters with the co-eval approximation (blue curves) begin to drop in amplitude at large scales. Further, the
amplitude of the drop increases for increasing $z$, while also moving to more intermediate $k$ scales. These trends are consistent with the 
results of \citet{Ghara:2015p7650}. The full redshift spans across these boxes grows
rapidly, from $\Delta z \sim 1.5$ to $\Delta z \sim 4.8$ at the edge of the HERA observing band. As a result, the $k$ modes within our fitting region are
now considerably more affected by the light-cone effect. This implies that the light-cone effects can be more severe on the 21cm PS during earlier 
epochs than the EoR, such as the X-ray heating epoch, when analysing the 21cm light-cone with equal co-moving distance cubes.

It is this increased contamination of the Fourier modes during the heating epoch that limits the constraining power on the heating parameters. In principle, 
the impact of this contamination could be reduced by considering alternative approaches to analyse the light-cone signature. For example, we could
reduce the observing bandwidth further at these higher redshift intervals to minimise the light-cone effects \citep[e.g.][]{Datta:2014p4990} or consider
more optimal statistical measures, such as the wavelet transform \citep[e.g.][]{Trott:2016p7921}.

From the figure, we also see that the lack of notable parameter recovery bias stems from the fact that the \fg{} model
is slow enough that the light-cone chunks sample all of the relevant astrophysical epochs well.
That is, the EoR, EoH and \lya{} peak are relatively well separated in time (redshift) \citep{Greig:2017p8496}.
Roughly speaking, the EoR spans from $z\simeq6-11$, the heating epoch from $z\simeq11-17$ and \lya{} (WF) coupling from $z\gtrsim17$. Therefore, the light-cone contains sufficient modes to constrain most of the astrophysical parameters, without notable bias.

Finally, it is illuminating to compare the ML ($\chi^{2}_{\rm min}$) values between the light-cone and co-eval simulations. When using the full light-cone, we 
obtain a $\chi^{2}_{\rm min} = 8.6$, whereas for the co-eval approximation we find $\chi^{2}_{\rm min} = 18.1$.
To objectively differentiate between the two scenarios, 
we use the Bayes information criterion (BIC; \citealt{Schwarz:1978p1,Liddle:2004p5730}),
\begin{eqnarray} \label{eq:BIC}
{\rm BIC} = -2\,{\rm ln}(\mathcal{L}) + k\,{\rm ln}(N),
\end{eqnarray}
where $\mathcal{L}$ is recovered from the $\chi^{2}_{\rm min}$ fit to the 21cm PS, $k$ is the number of free parameters 
and $N$ is the number of available data points. Taking the difference between the two BICs, we have $\Delta {\rm BIC} = 9.5$, which 
corresponds to strong evidence ($6 < \Delta {\rm BIC} < 10$) in favour of the light-cone model.
Therefore, despite the relatively similar constraints on the astrophysical parameter constraints, using co-eval cubes to analyse the full 
light-cone signal is still a relatively poor approximation.

\begin{figure*} 
	\begin{center}
		\includegraphics[trim = 0cm 0.5cm 0cm 0.5cm, scale = 0.9]{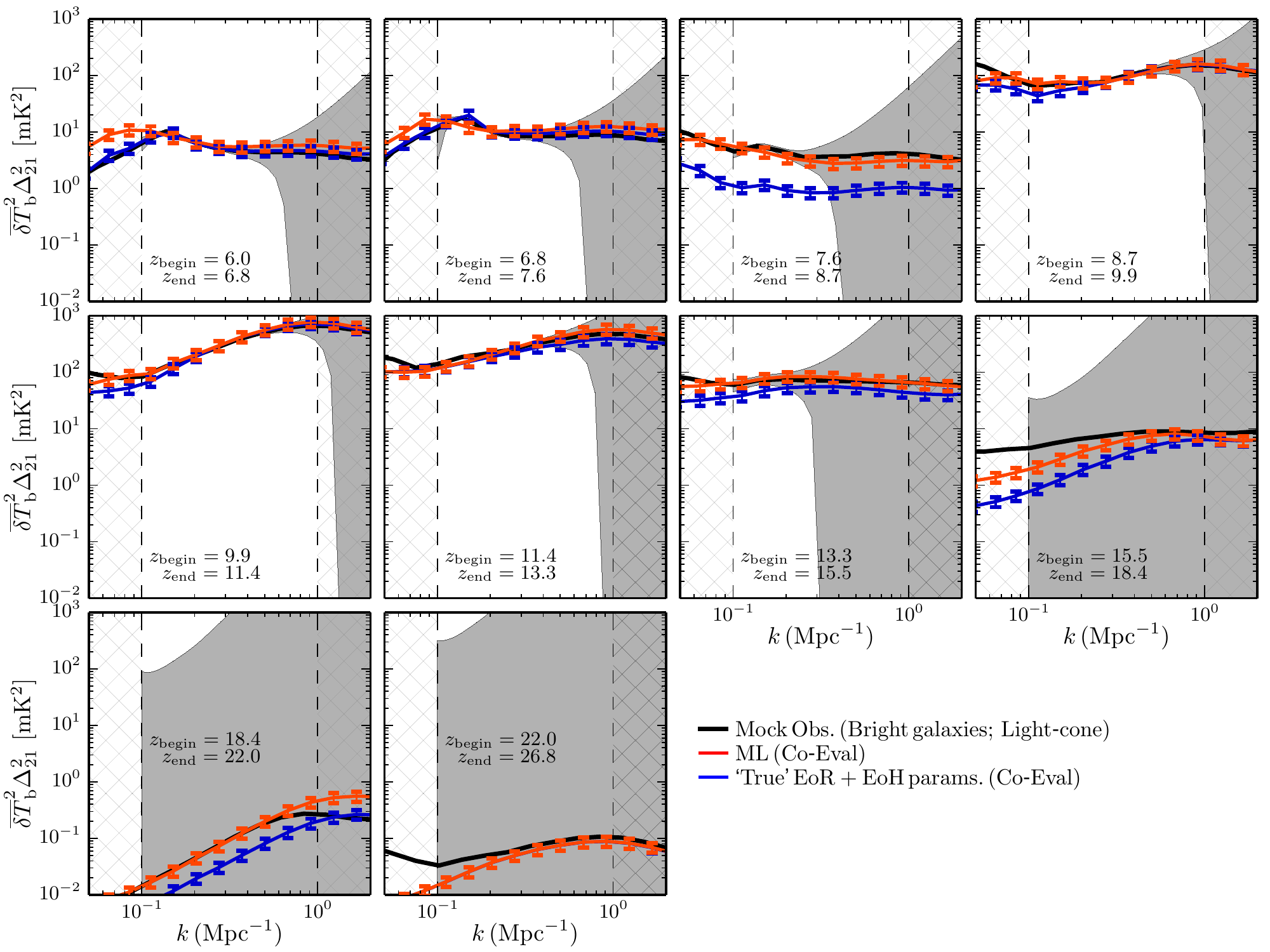}
	\end{center}
\caption[]
{The same as Figure~\ref{fig:BiasFaint}, except for our \textsc{bright galaxies} model. The fiducial \textsc{bright galaxies} parameter set corresponds to, ($\zeta$, $R_{\rm mfp}$, log$_{10}$$(T^{\rm min}_{\rm vir})$, log$_{10}$($L_{\rm X\,<\,2\,keV}/{\rm SFR}$), $E_{0}$, $\alpha_{X}$) = (200, 15, 5.48, 40.0, 0.5, 1.0).}
 \label{fig:BiasBright}
\end{figure*}

\subsection{Bright Galaxies}

Though the \fg{} model did not show any biases in the astrophysical parameters owing to assuming co-eval 21cm PS, it still
highlighted that the constraints could be significantly overestimated. The lack of a bias likely results from the relatively slow evolution of the
21cm signal. This should not be the case for the \bg{} model, where the EoR and heating epochs strongly overlap producing a more rapidly 
evolving 21cm signal, in effect maximising the potential bias from ignoring the light-cone effect.

In 
Figure~\ref{fig:LikelihoodBright}, we present the marginalised 1 and 2D posterior distributions when ignoring the light-cone effects (blue curves), 
and in the bottom row of Table~\ref{tab:BrightGalaxies} we provide the median and associated 16th and 84th percentiles for the recovered
astrophysical parameters. We note quite significant biases for all the astrophysical parameters for the \bg{} model\footnote{
Even $T^{\rm min}_{\rm vir}$ and $L_{\rm X\,<\,2\,keV}/{\rm SFR}$ are quite discrepant at high significance owing to their
relatively small errors.
}, in excess of $10\sigma$ away from the expected values. Given the large biases in the astrophysical parameters, it is less instructive to discuss
the relative percentage errors on their constraints.

In Figure~\ref{fig:BiasBright}, we compare the 21cm PS with and without the light-cone effect for the \bg{} model. The black curves,
correspond to the mock light-cone observation, the red curves the ML 21cm PS when ignoring the light-cone effect and the blue curve is the co-eval
21cm PS for the ``true'' \bg{} astrophysical parameters.

As the \bg{} model is driven by rare, biased sources, all epochs are strongly shifted towards lower redshift. In addition, to match the Planck constraints
on reionisation, an extremely large $\zeta$ was required. Combined, this results in the EoR strongly overlapping with the EoH.
Roughly speaking, the heating epoch commences at $z\sim12$ and continues until $z\sim8$, long after the EoR commences at around $z\sim10$. Therefore,
the first five equal co-moving distance light-cone cubes contain both the EoR and the heating epoch. Contrast this with the \fg{} model, where the first four redshift
intervals exclusively contained the EoR, and the second four the heating epoch. This highlights the rapidity of the 21cm signal evolution in the \bg{} model 
relative to the \fg{} model.

It is this rapid evolution that drives the strong bias in the astrophysical parameters. For example, consider the third panel in the top row of Figure~\ref{fig:BiasBright}.
For the \bg{} model in the co-eval approximation (blue curve), this redshift roughly corresponds to the brightness temperature contrast, $\delta T_{\rm b}$, approaching zero 
(transition between heating and EoR). However, the mock light-cone 21cm PS is an order of magnitude larger in amplitude. This difference arises as the light-cone
box picks up power from modes on either side of the transition, smoothing out the characteristic dip in amplitude that normally arises when 
$\delta T_{\rm b}$ approaches zero in the co-eval case. These large differences in the 21cm PS amplitude around the transitions between epochs with and
without the light-cone effect have been discussed previously in \citet{Ghara:2015p7650}.

In order for the co-eval approximation to instead match the mock observation, it must have this transition occurring between the two central redshifts 
of the light-cone cubes (i.e. between the co-eval redshifts). By significantly decreasing $\zeta$ (delaying EoR), this transition can be pushed to lower redshifts, 
enabling the co-eval 21cm PS to match the observed light-cone 21cm PS (red curve). 
However, as a consequence of pushing the transition to later redshifts, the co-eval approximation struggles to match the large scales of the 21cm PS 
at the lower redshifts (for example the ``knee'' in the 21cm PS around $k=0.15$~Mpc$^{-1}$). It 
adopts a larger $R_{\rm mfp}$ to compensate for the decreased $\zeta$, which pushes this ``knee'' to smaller $k$ (larger scales). 

As before, we can  quantify this by comparing the relative ML probabilities.
When directly MCMC sampling the light-cone, we recover 
$\chi^{2}_{\rm min} = 14.3$, which is slightly higher than for the \fg{} model. With the 21cm PS from the \bg{} typically having a higher amplitude owing to the
late and rapid evolution of the signal, the differences from the different assumed seeds are enhanced (note that the 21-cm cosmic variance is not strictly Gaussian as we assume here; e.g. \citealt{Mondal:2016p6091}),
likely causing the higher ML. If instead we consider the co-eval approximation
when fitting the \bg{} light-cone, we find $\chi^{2}_{\rm min} = 44.6$. This corresponds to a $\Delta {\rm BIC} \sim 30$, which corresponds to very strong evidence
($\Delta {\rm BIC} > 10$) against the co-eval approximation.

\section{Conclusion} \label{sec:summary}

Upcoming interferometers will detect the cosmic 21-cm signal over a wide range of redshifts and scales, providing a 3D map of the first billion years.  In order to more closely match the upcoming observations, here we extend our Bayesian MCMC EoR analysis tool, \cmmc{}, to sample full 3D light-cone realisations of the 21-cm signal on the fly.  The power spectra used to compute the likelihood, are now taken directly from the light-cone (instead of from co-eval cubes). Further enhancing realism, we include a treatment of RSDs, though they do not have a notable impact on the signal in our fiducial models.

With this new tool, we provide astrophysical parameter forecasts for a fiducial 1000 hour observation with HERA (constraints with SKA should be comparable; see \citealt{Greig:2017p8496}).  Consistent with our previous work which sampled co-eval cubes, we find that upcoming interferometers should be able to recover most EoR and EoH parameters at a precision of $\lsim1$--10 percent.  The exception to this is the X-ray spectral index, which is difficult to constrain independently of the bolometric soft-band luminosity.  This is exacerbated by light-cone sampling in the new \cmmc{}, since the evolution along the line-of-sight, ``smears-out'' the large-scale modes which can help constrain this EoH parameter.

We also quantify the biases in astrophysical parameter recovery which can stem from using power spectra constructed from co-eval boxes.  We show that the extent of such biases depend on how rapidly the 21-cm signal evolves with redshift.  Taking a \fg{} model as a mock observation, in which the EoR and EoH evolve relatively slowly and are relatively separated in redshift, we find no significant biases.  However for the \bg{} model, in which stars form efficiently only in massive, rapidly evolving halos, the recovered parameter constraints can be discrepant by $\gtrsim10\sigma$, if one ignores the light-cone.

However, we caution that
our setup was selected to be relatively optimistic, maximising  the potential co-eval bias. Compared to \citet{Greig:2017p8496}, we (i) decreased our modelling uncertainty from 20 to 10 per cent (ii) extended our fitting range to larger scales (from $k> 0.15 $ Mpc$^{-1}$ to $k> 0.1 $ Mpc$^{-1}$), assuming improved foreground removal.  Moreover, our likelihood statistic was the
spherically averaged 21cm PS, which is sub-optimal for a signal evolving along the line-of-sight. Therefore, when extracting astrophysical parameter 
constraints from the expected 21cm light-cone signal, ignoring the light-cone effect {\it can} have a significant affect on biasing the inferred 
astrophysical parameters. But, if careful, the co-eval approximation can do a reasonable job of modelling the signal.

Importantly this new, on-the-fly light-cone sampling brings the parameter extraction framework closer to the observations.  This means that the simulated signal can be more easily manipulated/corrupted to resemble real observations.  Moreover, it puts us in a position to realistically explore alternate statistics, including tomographical characterisation of the observed signal.
  
\section*{Acknowledgements}
This work was supported by the European Research Council (ERC) under the European Union's Horizon 2020 research and innovation programme (grant agreement No 638809 -- AIDA -- PI: Mesinger). The Centre for All-Sky Astrophysics in 3D (ASTRO 3D) is an Australian Research Council Centre of Excellence, funded by grant CE170100013.

\bibliography{21CMMC_wLC}

\end{document}